\documentclass[conference]{IEEEtran}
\IEEEoverridecommandlockouts

\usepackage{cite}
\usepackage{amsmath,amssymb,amsfonts}
 
\usepackage{graphicx}
\usepackage{textcomp}
\usepackage{xcolor}
\usepackage{multirow}
\usepackage{caption}

\usepackage{algpseudocode}
\usepackage{tabularray}
\usepackage[shortlabels]{enumitem}
 \usepackage{comment}
\usepackage{subcaption}

\def\BibTeX{{\rm B\kern-.05em{\sc i\kern-.025em b}\kern-.08em
    T\kern-.1667em\lower.7ex\hbox{E}\kern-.125emX}}

\usepackage{amsmath} 

\newcommand\Fig[1]{\textbf{Fig.}~\ref{#1}}
\newcommand\fig[1]{\textbf{Fig.}~\ref{#1}}

\newcommand\tab[1]{\textbf{Tab.}~\ref{#1}}
\newcommand\Equ[1]{\textbf{Eq.}~(\ref{#1})}
\newcommand\equ[1]{\textbf{Eq.}~(\ref{#1})}
 
 \newcommand\sect[1]{\textbf{Sec.}~\ref{#1}}

\begin{document}
 
\title{Bit Transition Reduction by Data Transmission Ordering in NoC-based DNN Accelerator  \thanks {The research has been supported in part by Vetenskapsrå (Swedish Research Council) through the LearnPower project (2020-03494).}
}

\author{
Yizhi Chen\IEEEauthorrefmark{1}, Jingwei Li\IEEEauthorrefmark{2}, Wenyao Zhu\IEEEauthorrefmark{1}, Zhonghai Lu\IEEEauthorrefmark{2}  \\
\IEEEauthorrefmark{1}KTH Royal Institute of Technology  \quad
\IEEEauthorrefmark{2}The Hong Kong University of Science and Technology (Guangzhou)  \\
\IEEEauthorrefmark{1}\{yizhic, wenyao\}@kth.se \quad
\IEEEauthorrefmark{2}ljli739@connect.hkust-gz.edu.cn, zhl@hkust-gz.edu.cn  
}

\maketitle

\begin{abstract}
 
\par  As Deep Neural Networks (DNN) are becoming essential, Network-on-Chip (NoC)-based DNN accelerators gained increasing popularity.   To save link power in NoC, many researchers focus on reducing the Bit Transition (BT).  We propose  '1'-bit count-based ordering method to reduce BT for DNN workloads. We provide a mathematical proof of the efficacy of proposed ordering. We evaluate our method through experiments without NoC and with NoC. Without NoC, our proposed ordering method achieves up to  20.38\%  BT reduction for floating-point-32 data and 55.71\%  for fixed-point-8 data, respectively. We propose two data ordering methods, affiliated-ordering  and separated-ordering to process weight and input jointly or individually and apply them to run full DNNs in NoC-based DNN accelerator. We evaluate our approaches under various configurations, including different DNN models such as LeNet and DarkNet, various NoC sizes with different numbers of memory controllers, random weights and trained weights, and different data precision.  Our approach efficiently reduces the link power by achieving up to  32.01\% BT reduction for floating-point-32 data and 40.85\% BT reduction for fixed-point-8 data.
 
\end{abstract}

\section{Introduction} 
Deep Neural Networks (DNN) have been widely deployed for extensive services. To efficiently run DNN models, DNN accelerators are essential. In recent years, Network-on-Chip (NoC)-based architectures \cite{shao2019simba,WENYAO2025nocdas} have gained significant popularity in designing DNN
accelerators.

The power of NoC is a critical constraint on performance as NoCs can consume a significant portion (10-30\%) of total chip power \cite{RLLinkBandwidthMWSCAS2020}.  Inside the NoC,  power is consumed by buffers, arbiters, switches, and links.  Links constitute a significant portion of the overall NoC's power \cite{kahng2009orion}. Links  consume nearly 17\% of the overall NoC power \cite{moulika2017data}, and Banerjee et al.  \cite{banerjeenocs2007power} report that link power accounts for around 42\% of the interconnect power.
 
Many techniques are proposed to mitigate link power consumption.    Paul \textit{et al.} \cite{ampadu2010adaptive} introduce a voltage control approach to lower link power.  Reza \cite{RLLinkBandwidthMWSCAS2020}  dynamically adjusts link bandwidths based on system utilization and application workload characteristics.  
Chen \textit{et al.} \cite{chen2015managing}  reduces the active link number. 
 
 \par Reducing bit transitions (BTs) is a popular research area aimed at reducing link power, focusing more on the data transmitted over links instead of link design itself.  BT is defined as a change from ’0’ to ’1’ or ’1’ to ’0’. Pagliari \textit{et al.} \cite{pagliari2017MP3Coding} extend the differential encoding with bounded approximations to achieve BT reduction. SILENT \cite{silentICCAD2004}  is proposed to reduce BT on serial links.   Sarman \textit{et al.} \cite{NICoding1BitVLSISOC2022} employ the Delta encoding technique to transform original data into a representation with fewer $'1'$ bits, thereby reducing BTs. As demonstrated by these works, the research community widely recognizes BT reduction as a key optimization target. 
\par  Mittal and Nag \cite{MITTAL2019Encodingsurvey}  note that most existing BT reduction techniques have been proposed for CPUs. Given the vastly different architectures and design philosophies between CPUs and AI accelerators, Mittal and Nag \cite{MITTAL2019Encodingsurvey} emphasize the need for developing accelerator-specific techniques.  Targeting BT reduction in NoC-based DNN accelerators (NOC-DNA), we conclude our contribution as follows:
\begin{enumerate}
 \item \textbf{Mathematical analysis}
     We provide a mathematical proof of the efficacy of our '1'-bit count-based descending ordering approach in reducing BTs.
  \item \textbf{Effective ordering for DNN workloads:} Our approach achieves up to 40.85\% BT reduction when running full DNN workload in NoC, and we utilize the inherent order-insensitivity in DNN convolutional and linear layers. %
 
 \item \textbf{ Validation across various configurations:} 
  We comprehensively validate our method through: without NoC and with NoC, different NoC sizes, different DNN models, and different data formats. 
\end{enumerate}

\par The rest of the paper is organized as follows: \sect{sec:related_work} covers the related work. \sect{sec:problem} provides problem definitions and a solution with mathematical proof. \sect{sec:results} presents the experimental results and \sect{sec:conclusion} concludes the paper.
\section{Related work}
\label{sec:related_work}
The most straightforward way to reduce link power is to adjust the link design.  Mineo \textit{et al.} \cite{mineo2013runtime} introduce that the voltage swing in links significantly impacts power consumption.  Paul \textit{et al.} \cite{ampadu2010adaptive} control the voltage to reduce link power.   Configurable architecture is also used to reduce link power. Reza \cite{RLLinkBandwidthMWSCAS2020} utilizes reinforcement learning to dynamically decide the link bandwidth.  Chen \textit{et al.} \cite{chen2015managing} reduce the active link number by deactivating associated L2 cache banks.

Reducing BT absorbs researchers' interest \cite{pagliari2017MP3Coding}, and  BT reduction rates vary for different designs and applications.  Silent \cite{silentICCAD2004} proposed a serialized low-energy transmission coding for NoC and reduced switching activities by 32.35\%. Stan \textit{et al.} \cite{1995fullbusinvert} reduce the average number BT by 25\% compared to unencoded data. But it needs extra lines on bus.   Pagliari  \textit{ et al.} \cite{pagliari2017MP3Coding} achieve bit reduction rate from 42.73\% for transmitting document files (.doc). 
 Ghosh \textit{ et al.} \cite{ghosh2014dataCorelationCoding} proposed a method utilizing the correlations in the data to reduce self-transitions and achieve an average reduction of 25\%. These two methods require prior information about the whole dataset.

Firstly, our method is not a bus-encoding method and operates without additional links. It also requires no prior information about the whole dataset.
Secondly, while most existing coding approaches are designed for CPUs \cite{MITTAL2019Encodingsurvey}, our ordering methods are designed for DNN workload.

\section{BT reduction problem and mathematical solution} \label{sec:problem}
\subsection{Problem definition}
 
\par  In  NoCs,  BTs occur when consecutive flits with differing bit patterns ('0'→'1' or '1'→'0') traverse the links.   The problem is defined as \textbf{minimizing the expected number of BTs between two consecutive flits }.
\par We first consider comparing two 32-bit  numbers passing the same 32-bit links. 
Assuming the first number contains $x$-bits of $\text{'1'}$, and the second number contains $y$-bits of $\text{'1'}$, where $x$ and $y$ are non-negative integers between 0 and 32, \Equ{e.p1} shows the probability of bit transition on each single-bit link.
\begin{equation} \label{e.p1}
  \begin{aligned}
P(t_{OneLink}) & = 1 - P(both \ '0') - P(both \ '1') \\
& = 1 - \frac{(32-x)(32-y)}{32\times 32} - \frac{xy}{32\times 32}
  \end{aligned}
\end{equation}

\begin{figure}[htbp] %
	\centering 
\includegraphics[trim=0.0cm 0.4cm 0.0cm 0.53cm, clip, width=0.4\textwidth]{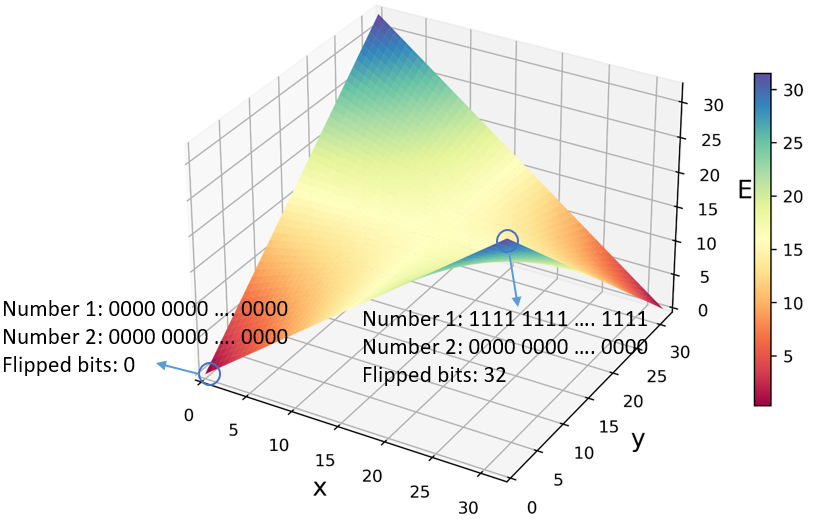} 
	\caption{Expectation of BT on two 32-bit numbers.}
	\label{twonumberplot}
\end{figure}

Assuming different bits are independent and identically distributed, the expectation of the BT is \equ{e.p2}. We present expectations of BT across different $\text{x}$ and  $\text{y}$  in \Fig{twonumberplot}.  
\begin{equation} \label{e.p2}
E = 32 \times P(t) = x + y - \frac{xy}{16}, 
\end{equation}
 
Then we expand from comparing two 32-bit numbers to comparing two flits, containing a series of numbers. Assuming each flit contains $N$ 32-bit numbers, $x_i$ and $y_i$ representing the count of $'1'$ bit in $i$th number in flit $f_1$ and flit  $f_2$, respectively ($1 \leq i \leq N$). The total expectation of BTs is 
\begin{equation} \label{e.p3}
E_t = \sum_{i=1}^{N} E_i = \sum_{i=1}^{N} x_i + \sum_{i=1}^{N} y_i - \frac{ \sum_{i=1}^{N} (x_i \times y_i)}{16}
\end{equation}
 
Since the data is determined by the application and is therefore fixed, the multisets of $x$ and $y$ are also fixed. 
  Hence, regardless of changing positions of 32-bit  numbers, $ \sum_{i=1}^{N} x_i + \sum_{i=1}^{N} y_i$  is constant.  The goal of minimizing $E_t$ is to find an arrangement solution for $x$ and $y$ to maximize
\begin{equation} \label{e.p4}
F = \sum (x_i \times y_i).
\end{equation}. 
\vspace{-3mm}
\subsection{ '1'-bit count-based optimal strategy}
\subsubsection{Count-based strategy}
\noindent The intuitive way is to go through all options to  find the best  $F$ in \equ{e.p4}. However, for inserting 2N values in two N-value flits, there are $(2N)!$ options. For  128-bit link with fixed-8 data, we have overall $32!$ options, which is larger than $2.6 \times 10^{35}$ options. Instead of exhaustive search, we propose a count-based ordering to maximize $F$ in \equ{e.p4} as described below.  
\begin{enumerate}[(1)]
    \item Given two N-number flits, $f_1$ and $f_2$, the values of the numbers are    $[V_{1_{oldf1}};V_{2_{oldf1}};...V_{N_{oldf1}}]$ in $f_1$ 
    and $[V_{1_{oldf2}};V_{2_{oldf2}};...V_{N_{oldf2}}]$ in $f_2$. The '1'-bit counts of each number in $f_1$ and $f_2$ represented as $[x_{1_{old}};x_{2_{old}};...x_{N_{old}}]$ and $[y_{1_{old}};y_{2_{old}};...y_{N_{old}}]$, respectively. The initial order of both the $x$ and $y$ series is arbitrary—for instance,   $[x_{2_{old}}>x_{3_{old}}>y_{1_{old}}...]$, with no inherent ordering. %
    \item After applying count-based ordering, we rearrange the values to be  $[V_{1_{f1}};V_{2_{f1}};...V_{N_{f1}}]$ and $[V_{1_{f2}};V_{2_{f2}};...V_{N_{f2}}]$ to ensure  a strictly decreasing sequence of '1'bit counts: $[x_1> y_1>x_2>y_2...>x_N>y_N]$.  %
\end{enumerate}
 Our transformation only changes the order of each number $V$ and maintains strict numerical equivalence between the original and reordered sets. For example, we are just changing the numbers of previous flits from $[V_A;V_B]$ in $f1$ and $[V_C;V_D]$ in $f2$  to be $[V_A;V_C]$  in $f1$  and $[V_B;V_D]$ in $f2$.
  
\subsubsection{Proof with inductive steps}
\noindent We provide mathematical proofs with inductive steps to show that this ordering is a globally optimal ordering.  Given two flits carrying  N-element sequences and the series of'1'-bit counts are  $X$ and $Y$, the initial state is that the values in $X$ and $Y$ are randomly placed:
\begin{enumerate}[(1)]
\item Local Pairwise Optimization: For any two number  $[V_{i_{oldf1}};V_{j_{oldf1}}]$  in first flit and two numbers $[V_{i_{oldf2}};V_{j_{oldf2}}]$ in flit 2, where $i<j$, let their initial bit-counts be  $[x_{i_{old}},x_{j_{old}}]$ in flit $f1$ and $[y_{i_{old}},y_{j_{old}}]$ in  flit $f2$. Enforcing the ordering constraint  $[x_i >y_i> x_j>y_j]$ maximizes the pairwise product sum:
   $(x_i*y_i+x_j*y_j)>= (x_{i_{old}}* y_{i_{old}}+x_{j_{old}}*y_{j_{old}})$. This inequality holds trivially and can be easily verified through exhaustive enumeration, as there are only four elements.  Hence, to maximize $\sum xy$ in \equ{e.p4} for these four numbers, we should order the values to be $V_{i_{f1}};V_{j_{f1}}$ and$V_{i_{f2}};V_{j_{f2}}$  with corresponding  '1'-bit count being $[x_i;x_j ]$ and $[y_i;y_j]$, where $[x_1 >y_1> x_2>y_2]$. 

\item Global Optimization: Iteratively applying this rule to all pairs $(i,j)$ until convergence yields a globally optimal ordering: $[x_1> y_1>x_2>y_2...>x_N>y_N]$.
\end{enumerate} 
 
As proof, we demonstrate the ordering's optimality;  the choice of sorting algorithms (Bubble Sort/Bitonic Sort/Merge Sort) to achieve the ordering is not discussed.  This proof provides clear and essential guidance in two-flit scenarios; we evaluate our methods later with the complex interleaving of flits in NoC carrying real DNN workloads.

\section{Data transmission ordering for NOC-DNA}\label{method}
 
\begin{figure}[htbp] %
	\centering 
\includegraphics[ trim=0.5cm 0.4cm 0.0cm 0.2cm, clip, width=9cm ]{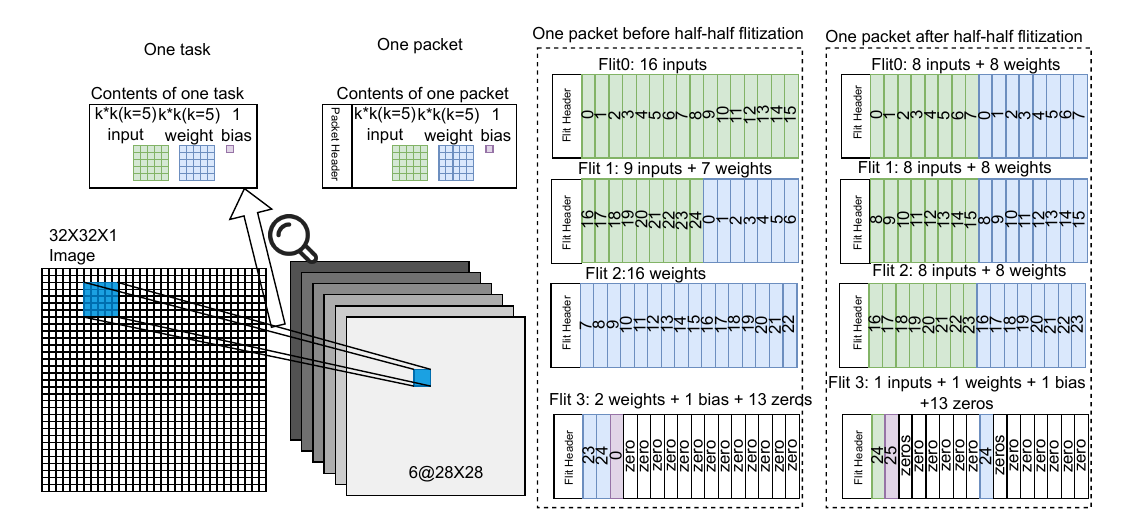} 
	\caption{Dividing inputs and weights into flits.}
	\label{half and half}
\end{figure}

A typical neuron calculation in NOC-DNA involves the inputs and weights. 
By dividing the inputs into the left half and the weights into the right half of the flit as shown in the right side of  \fig{half and half}, we can focus solely on the bit comparison between the weights part or on the whole flits. 

\begin{figure}[htbp] %
    \centering %

    \begin{subfigure}[b]{0.46\textwidth} %
        \centering
        \includegraphics[trim=0.44cm 0.2cm 0.4cm 0.43cm, clip,width=\linewidth]{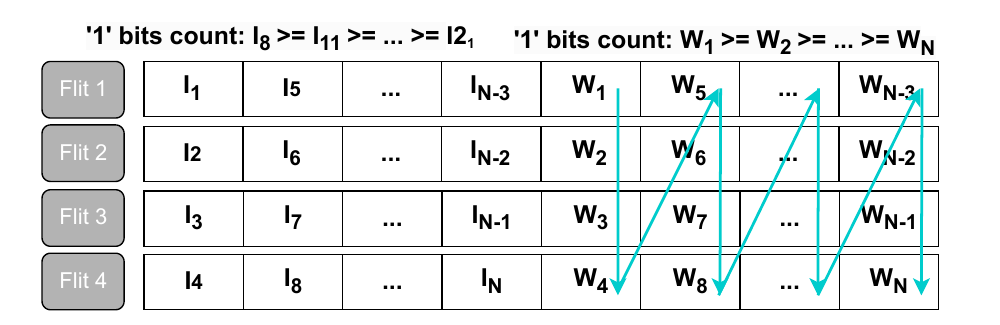} 
       	  \caption{Principle of affiliated-ordering.}
	\label{togetherordering}
    \end{subfigure}

    \vspace{2mm}
    \begin{subfigure}[b]{0.46\textwidth} %
        \centering 
        \includegraphics[trim=0.42cm 0.2cm 0.4cm 0.43cm, clip,width=\linewidth]{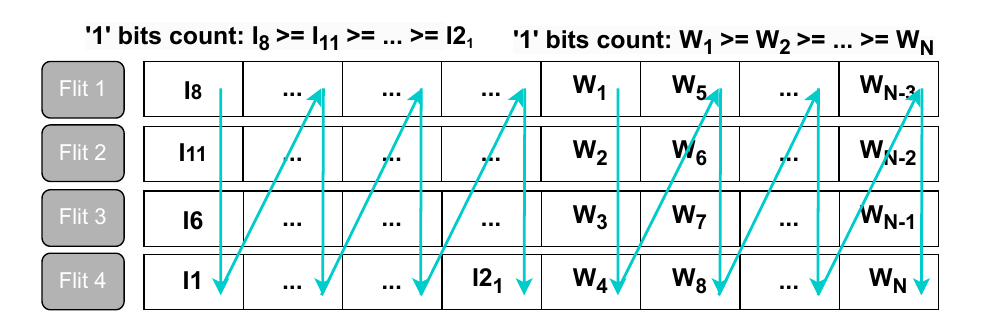} 
       \caption{Principle of separated-ordering.}
	\label{coreseperateordering.pdf}
    \end{subfigure}
    \caption{Affiliated-ordering (a)  and separated-ordering (b).}

\end{figure}

We present the basic idea of arranging data to reduce BTs in \Fig{togetherordering}. The weights are ordered following their own descending order of '1'-bit count, highlighted by the arrows.  According to different ways to process inputs and weights, we propose two methods: 
\begin{itemize}
    \item Affiliated-ordering: weights, together with inputs, are ordered based on the '1'-bit counts of weights.
    \item Separated-ordering: weights and inputs are ordered according to their own '1'-bit counts, respectively.
\end{itemize}

\subsection{Affiliated-ordering}
\label{weightbasedAndSeperateOrdering}
Affiliated-ordering arranges the weights in descending order based on their '1'-bit count. Simultaneously, the inputs do not follow their own ordering trend, and no arrows for inputs. Instead, inputs remain affiliated with corresponding weights.

\begin{figure}[htbp] %
	\centering 
\includegraphics[trim=0.64cm 0.4cm 0.4cm 0.62cm, clip, width=0.48\textwidth ] {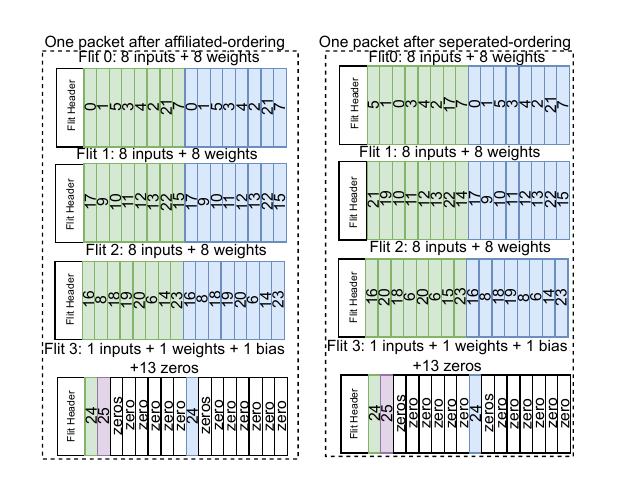} 
	\caption{Affiliated-ordering vs separated-ordering.}
	\label{ExtraID}
\end{figure}

In \fig{ExtraID}, we do not order the padded zeros to highlight the ordering of inputs and weights. Before affiliated-ordering, original inputs and weights are randomly placed. After ordering based on '1'-bit counts, we change the positions of each weight, and the indices are out of order.  Crucially, the inputs are changed,  but the indices of inputs and weights in corresponding positions are kept paired. The disadvantage is that it only gains BT reduction from the weights' part, and the advantage is that it maintains the paired relationship between weights and inputs.

\subsection{Separated-ordering}\label{weightbasedAndSeperateOrdering} 

In separated-ordering, as illustrated in \fig{coreseperateordering.pdf}, the key difference is that the inputs are independently placed according to their own order based on '1'-bit count (highlighted by arrows), instead of pairing with the weights.

In \fig{ExtraID}, the indices of inputs and weights in corresponding positions are not the same, meaning that the inputs and weights are not paired.   This separated-ordering method reduces not only the weight part's BT but also the input part's BT.

\subsection{Advantages related to DNN and NoC}
\subsubsection{Low overhead for recovering data}
Conventional data encoding schemes require both encoding and decoding steps. Our data transmission ordering method not only reduces BT but also keeps the values' integrity, which can allow no decoding process in NOC-DNA. 

\begin{figure}[htbp] %
	\centering 
\includegraphics[width=0.49\textwidth]{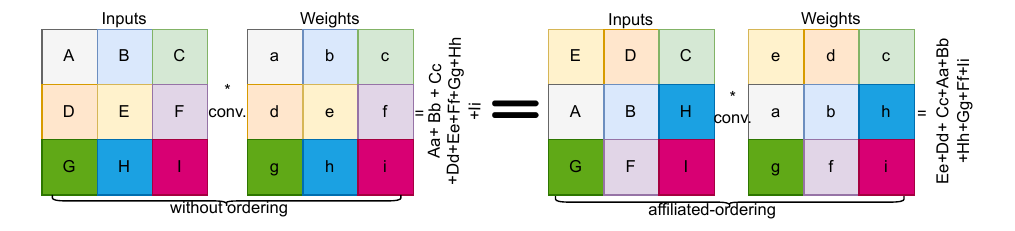} 
	\caption{3$\times$3 convolution before and after affiliated-ordering.}
	\label{noextraID.pdf}
\end{figure}

For affiliated-ordering, only the ordering operation is performed. No subsequent deordering is required because the convolutional and linear layers in DNNs exhibit order invariance: the output remains unchanged when reordering paired weights and inputs. One example is shown in \fig{noextraID.pdf}.     As long as the input and weight are paired, the output result will not be changed and is equal to the sum of the products of corresponding inputs and weights, $A\times a+B\times b+ \cdots + I\times i$.  Since affiliated-ordering  preserves  the pairing relationship between inputs and weights, we don't need to recover the data.

For separated-ordering, just a minimal-bit-width index is required, and this costs less than complex bit-wise operations for traditional decoding methods.

\subsubsection{Near off-chip memory placement }
We place the ordering unit near off-chip memory instead of inside the NoC to sort the data. The off-chip resource constraints are less stringent, and Li \textit{et al.}  \cite{li2020imc} have proposed modern technology for in-memory sorting. Additionally, our ordering unit's count is the same as the number of memory controllers (MCs) and much smaller than the number of routers. 
\begin{figure}[h] %
	\centering 
\includegraphics[trim=0.64cm 0.4cm 0.4cm 0.43cm, clip,width=0.43\textwidth]{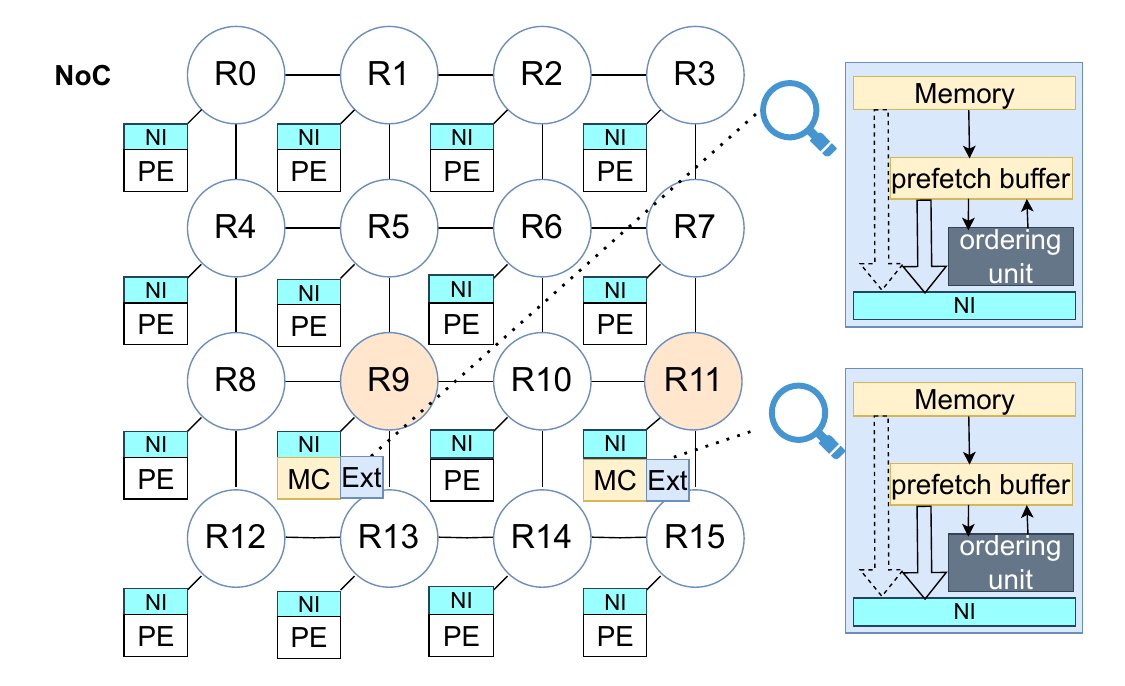} 
	\caption{Ordering unit's placement.}
	\label{nearmemory.pdf}
\end{figure} 
\subsubsection{Hide extra latency} In DNN accelerators, the outputs computed at each layer are not immediately consumed by the next layer, as there are still many tasks of the current layer pending.  This creates a sufficient time window for ordering data before they serve as the next layer's inputs.  The layer-level interval effectively hides ordering latency from the critical path.

\section{Experiment results} \label{sec:results}
\par
\begin{figure}[htbp] %
	\centering 
\includegraphics[ width= 0.5\textwidth]{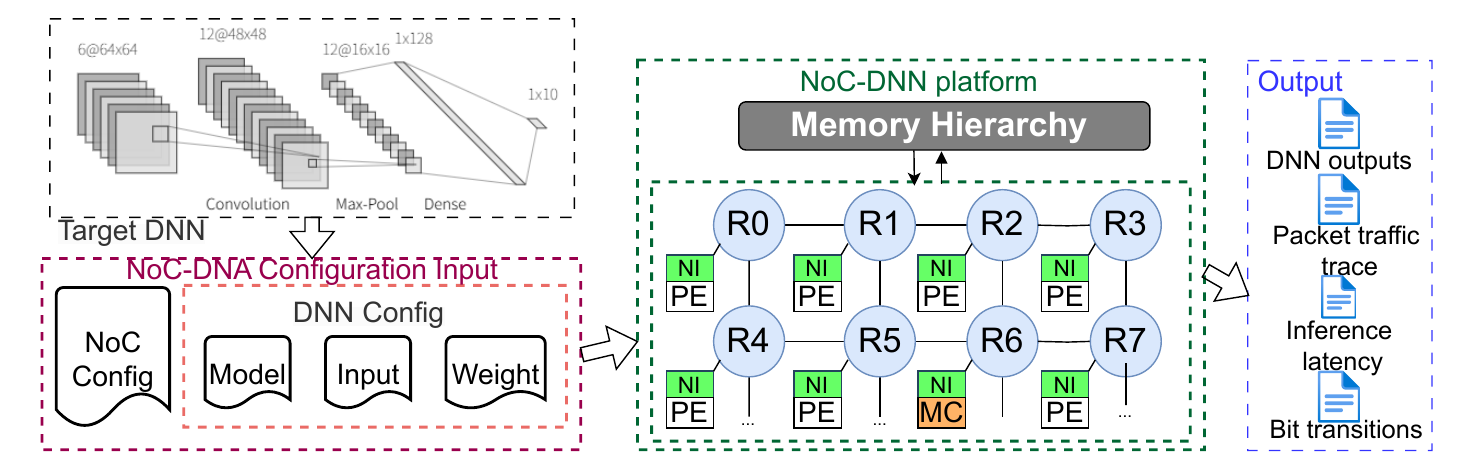} 
\caption{Overview of the used NOC-DNA.}
	\label{nocdasoverview.pdf}
\end{figure}

 First, we evaluate the ordering technique without NoC to show the effectiveness in simple scenarios.  Then we evaluate our ordering method on NocDAS \cite{WENYAO2025nocdas}, a NOC-DNA,  where flits carry actual DNN data and undergo complex and real router-to-router transmission as shown in \fig{nocdasoverview.pdf}.

 \par
\begin{figure}[htbp] %
	\centering 
\includegraphics[width=0.38\textwidth]{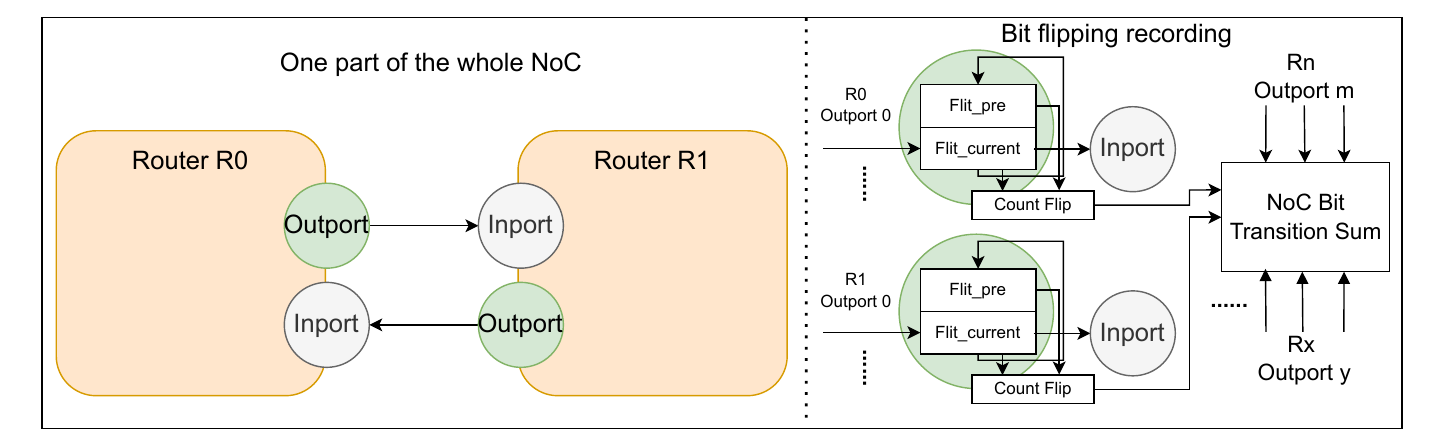} 
	\caption{Recording BT  for performance evaluation.}
	\label{outportFlip.pdf}
\end{figure}

 Our designed BT recording method is shown in \fig{outportFlip.pdf}. The current flit and the previous are compared, and BTs are accumulated across the entire NoC. BT recording is solely for performance evaluation, and the  flit storage and BT summation should not be considered overheads.   

\subsection{BT reductions without NoC}

\par  We generate 10,000 packets without NoC and measure the BTs of random comparisons between flits. Real weights are used as the payloads of flits, including trained LeNet weights and randomly initialized weights. Zeros are padded when the weight's kernel size doesn't exactly match the flit size. Two common data types, 32-bit floating-point (float-32) and 8-bit fixed-point (fixed-8), are used.

\par
\begin{figure}[htbp]
    \centering
    \begin{minipage}[b]{0.24\textwidth}
       \centering 
\includegraphics[trim=2.3cm 3cm 5.5cm 2cm,clip,width=\textwidth]{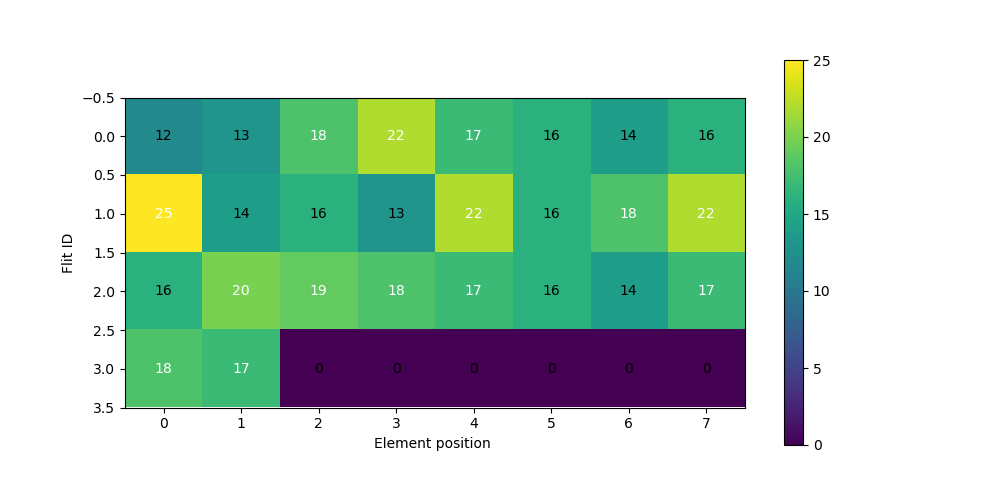} 
	\label{heatmapbefore.png}
    \end{minipage}
    \begin{minipage}[b]{0.24\textwidth}
       	\centering 
\includegraphics[trim=2.3cm 3cm 5.5cm 2cm,clip,width=\textwidth ]{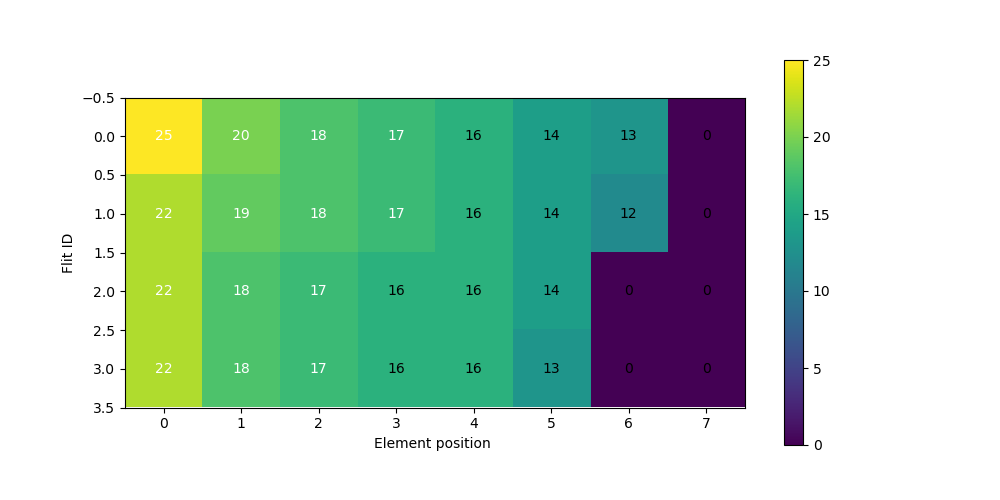} 
	\label{heatmapColmajorafter.png}
    \end{minipage}
    \caption{Data before ordering (Left) and after ordering (Right).}
    \label{heatmapordering}
\end{figure}

 We present one example in \fig{heatmapordering}. The y-axis is the flit ID and eight squares in each row represent eight weights in one flit, while the number inside the square indicates the  '1'-bit count of each weight. The data are descending ordered by their '1'-bit count, as shown in the right of \fig{heatmapordering}.
 
\subsubsection{Summary of BT reduction without NoC}

\par  We summarize the average BT reductions in \tab{pythoncollsionsummary}.   Proposed data transmission ordering method achieves a BT reduction of 20.38\% for float-32 random weights and 18.92\% for float-32 trained weights.  For fixed-8 weights, the impact of ordering is more significant and the reductions are 27.70\% and 55.71\%. %

\begin{table}[!htbp]
\centering
\caption{BT reduction without NoC}
\label{pythoncollsionsummary}
\begin{tabular}{|c|c|cc|c|}
\hline
\multirow{2}{*}{Weights}                                   & \multirow{2}{*}{Flit size (bit)} & \multicolumn{2}{c|}{BTs per flit}                                                                                                        & \multirow{2}{*}{\shortstack{BT Reduction\\rate}} \\ \cline{3-4}
                                                           &                       & \multicolumn{1}{c|}{\begin{tabular}[c]{@{}c@{}}Baseline\end{tabular}} & \begin{tabular}[c]{@{}c@{}}Ordered\end{tabular} &                            \\ \hline
\begin{tabular}[c]
{@{}c@{}}Float-32 random\end{tabular}  & 32$\times$8                  & \multicolumn{1}{c|}{113.27}                                                      & 90.18                                                   & 20.38\%                    \\ \hline
\begin{tabular}[c]{@{}c@{}}Fixed-8 random\end{tabular}     & 8$\times$8                   & \multicolumn{1}{c|}{31.01}                                                       & 22.42                                                   & 27.70\%                    \\ \hline
\begin{tabular}[c]{@{}c@{}}Float-32 trained\end{tabular} & 32$\times$8                  & \multicolumn{1}{c|}{112.80}                                                      & 91.46                                                   & 18.92\%                    \\ \hline
\begin{tabular}[c]{@{}c@{}}Fixed-8 trained\end{tabular}    & 8$\times$8                   & \multicolumn{1}{c|}{30.55}                                                       & 13.73                                                   & 55.71\%                    \\ \hline
\end{tabular}
\end{table}

\subsubsection{Bit distribution and transition probability}
The value distribution of weights has been widely studied; however, the bit-level distribution of values remains an interesting topic. 

\par
\begin{figure}[htbp]
    \centering
    \begin{minipage}[b]{0.24\textwidth}
    \includegraphics[trim=1.5cm 0cm 1.5cm 2.2cm, clip,width=\textwidth]{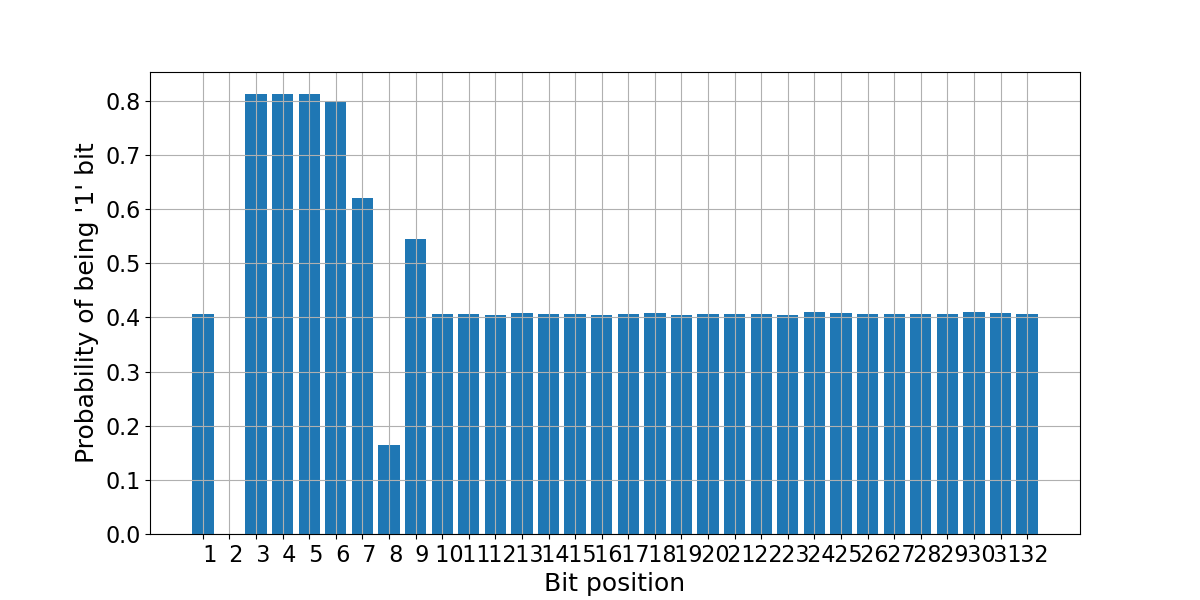}
\label{32FloatingBeingOne_26Of4flits_randomInitialize.png}
    \end{minipage}
    \begin{minipage}[b]{0.24\textwidth}
      \includegraphics[trim=1.5cm 0cm 1.5cm 2.2cm, clip,width=\textwidth]{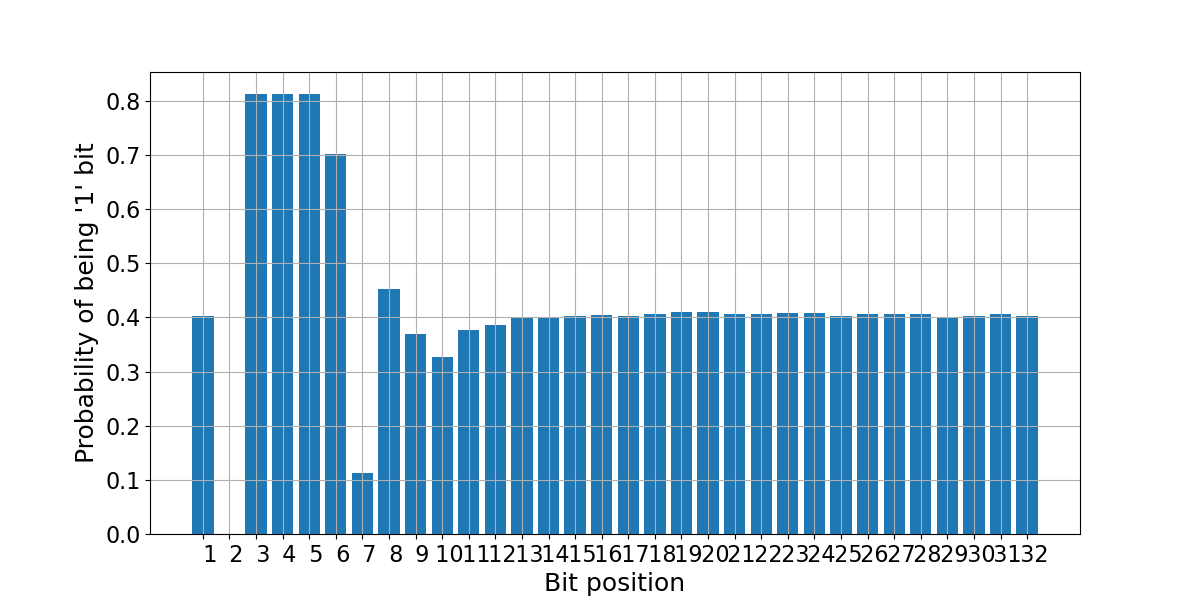}
\label{32FloatingBeingOne_26Of4flits_readLeNet.png}
    \end{minipage}

 \vspace{1mm}
    \begin{minipage}[b]{0.24\textwidth}
        \includegraphics[trim=1.5cm 0cm 1.5cm 2.2cm, clip, width=\textwidth]{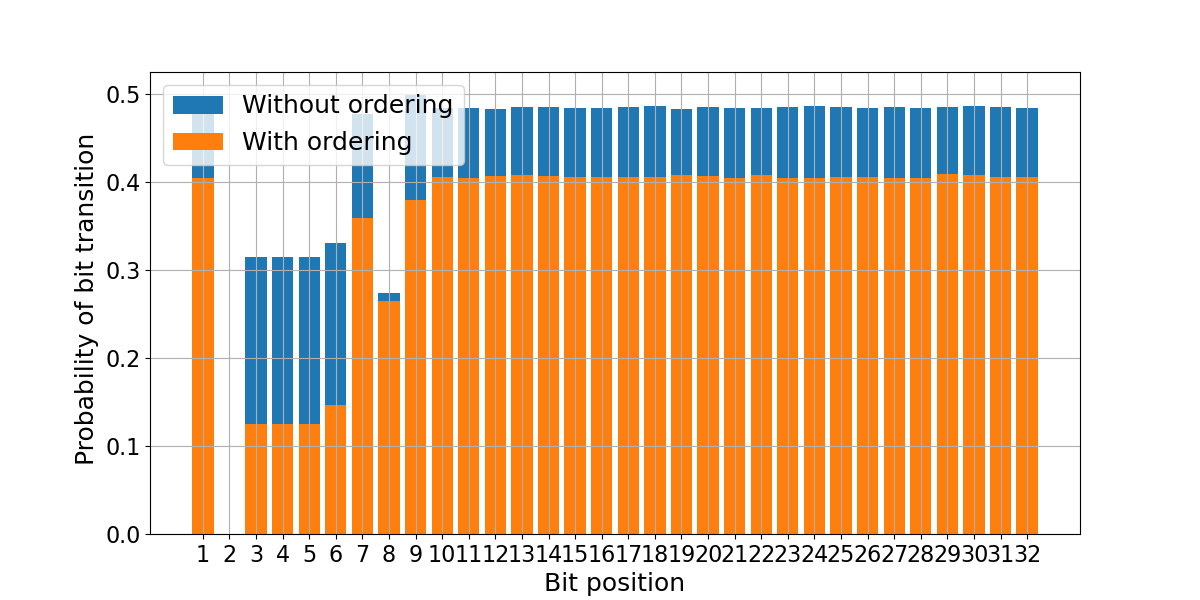}
\label{32ExistFlipping_26Of4flits_randomInitialize.png}
    \end{minipage}
    \begin{minipage}[b]{0.24\textwidth}
   \includegraphics[trim=1.5cm 0cm 1.5cm 2.2cm, clip, width=\textwidth]{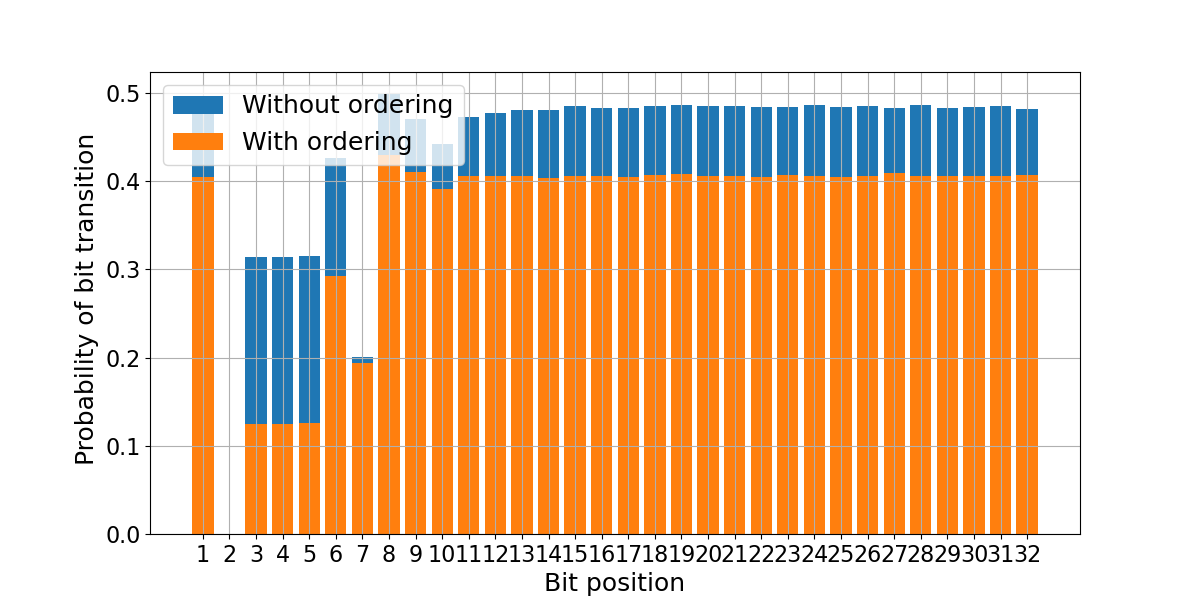}
    \label{32ExistFlipping_26Of4flits_readLeNet.png}
    \end{minipage}
    
    \caption{Analyzing different floating-point-32 weights: probability of "1"-bit at each bit position for random weights (top-left) and  LeNet trained weights (top-right);  probability of transition occurring at each bit position for random weights (bottom-left) and  LeNet trained weights (bottom-right).}
    \label{FloatPython}
\end{figure}

 For float-32 weights, we present the bit distribution across the 32  positions at the top of \fig{FloatPython}. We can obviously distinguish the sign bit, 8-bit exponent, and 23-bit mantissa as they have different patterns. The first sign bit is around 0.5. For the exponent part, we observe that the beginning bits (2-6 in the x-axis) and the last three bits (7-9 in the x-axis) show notable differences.   For the mantissa part (beyond position 9 on the x-axis), the results reveal that random data exhibits more uniform bit distribution, whereas trained data displays some fluctuations in tail bits. 

As for BT probability at the bottom in \fig{FloatPython}, the orange color is lower than the blue color, showing that the ordering method makes the transition probability lower than without ordering. 

\par
\begin{figure}[htbp]
    \centering
    \begin{minipage}[b]{0.24\textwidth}
    \includegraphics[trim=1.25cm 0cm 0cm 1.2cm, clip,width=\textwidth]{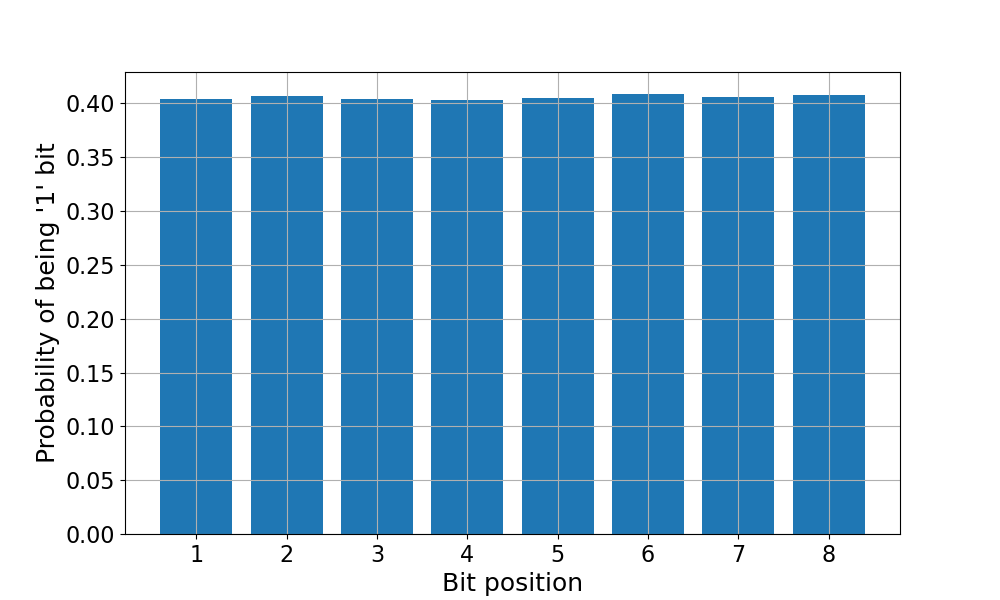} 
 \label{randomfixedBeingOne.png}
    \end{minipage}
    \begin{minipage}[b]{0.24\textwidth}
      \includegraphics[trim=1.25cm 0cm 0cm 1.2cm, clip,width=\textwidth]{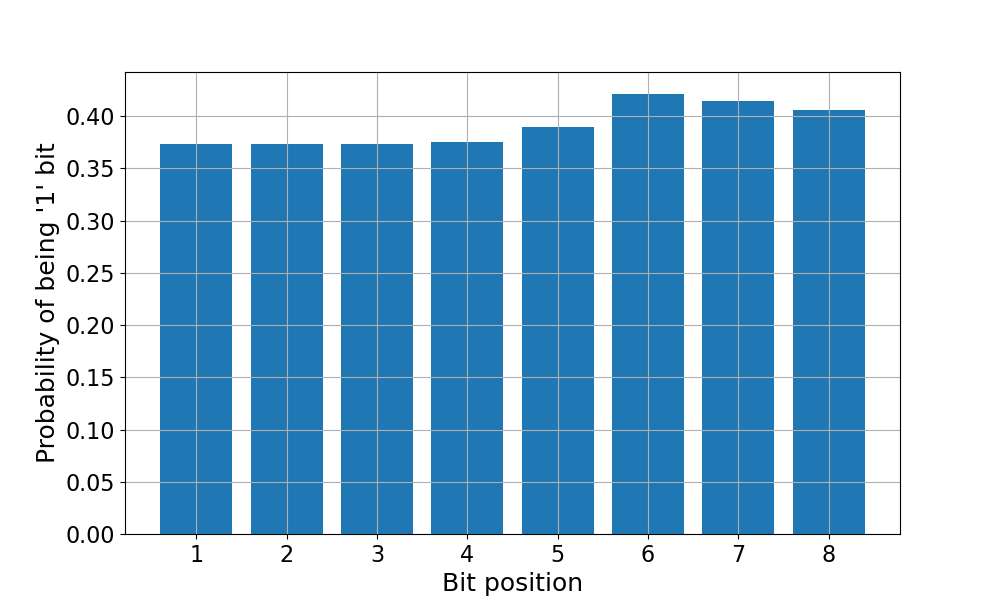} 
	\label{trainedFixedBeingOne.png}
    \end{minipage}

 \vspace{-1mm}
    \begin{minipage}[b]{0.24\textwidth}
        \includegraphics[trim=1.5cm 0cm 0cm 1.2cm, clip,width=\textwidth]{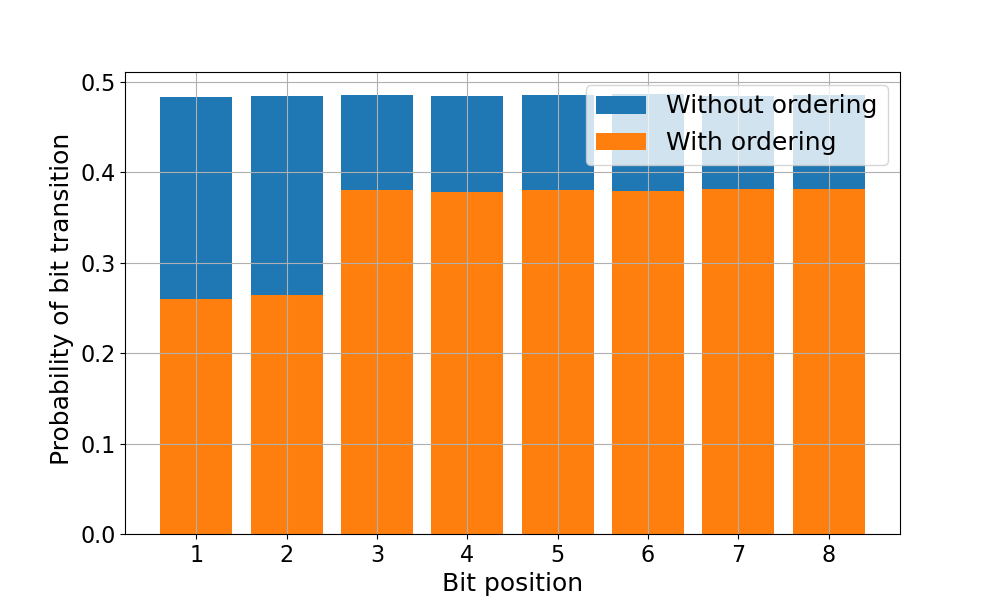}
        \label{RandomfixedTransition.png}
    \end{minipage}
    \begin{minipage}[b]{0.24\textwidth}
        \includegraphics[trim=1.5cm 0cm 0cm 1.2cm, clip,width=\textwidth]{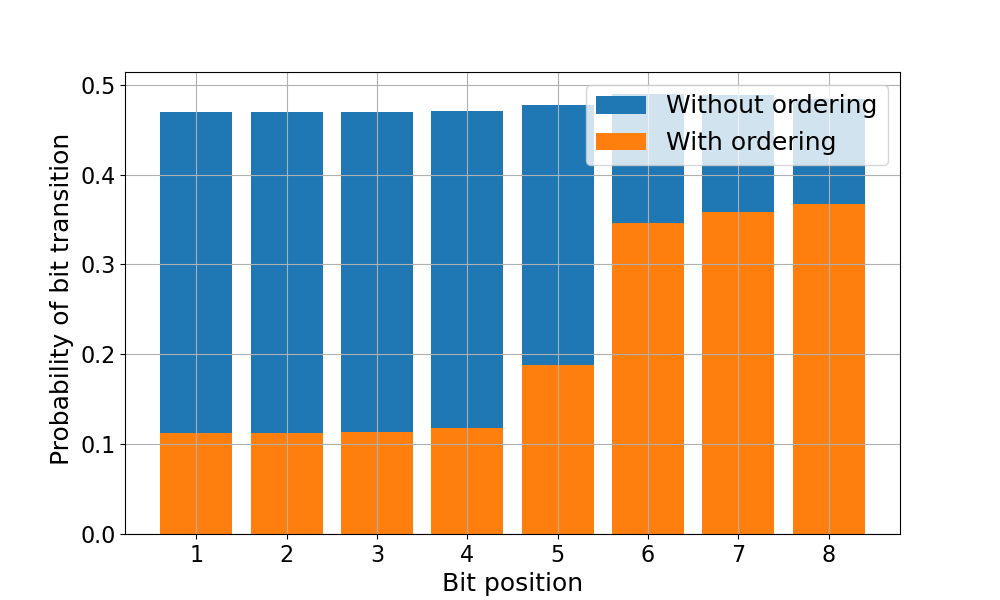}
\label{trainedFixedTransition.png}
    \end{minipage}
       \caption{Analyzing different fixed-8 weights: probability of "1"-bit at each bit position for random weights (top-left) and  LeNet trained weights (top-right);  probability of transition occurring at each bit position for random weights (bottom-left) and LeNet trained weights (bottom-right).}
        \label{Fixed point python results.}
\end{figure}

  For BT of fixed-8 data, as shown in \fig{Fixed point python results.}, the impact of ordering is more significant. In the bottom of \fig{Fixed point python results.}, the gap between the orange and blue bars  
 is significant, particularly for trained fixed-point 8-bit data. The distinct gap in the bottom-right of  \fig{Fixed point python results.} align with the highest BT reduction rate 55.71\% in \tab{pythoncollsionsummary} for fixed-8 data.

\subsection{BT reduction with real traffic in NoC}
We conducted experiments to run a complete DNN on a NOC-DNA \cite{WENYAO2025nocdas}, where numerous real packets are injected and transferred with interleaving of flits from different packets.  The NoC use X-Y routing, 4 virtual channels (VCs) with a 4-flit-depth buffer per VC. Dual data formats and two link bandwidths are used: 512-bit link for 16 float-32 values, and 128-bit link for 16 fixed-8 values.  
 \begin{itemize}
   \item We evaluate three configurations: baseline without ordering (O0), affiliated-ordering (O1), and separated-ordering (O2) as introduced in \sect{weightbasedAndSeperateOrdering}. 
     \item We explore different NoC sizes, including a default 4$\times$4 NoC with 2 memory controllers (MCs) (MC2), 8$\times$8 NoC with 4 MCs (MC4), and 8$\times$8 NoC with 8 MCs (MC8).
    \item  We run experiments on different DNN models, including LeNet and a DarkNet-like model. We reduce the input size for DarkNet to $64\times64\times3$ to speed up the simulation.  
  \end{itemize}

\subsubsection{Different NoC sizes}
\par
\begin{figure}[htbp] %
	\centering 
\includegraphics[trim=0.64cm 0.4cm 0.4cm 0.63cm, clip,width=0.48\textwidth]{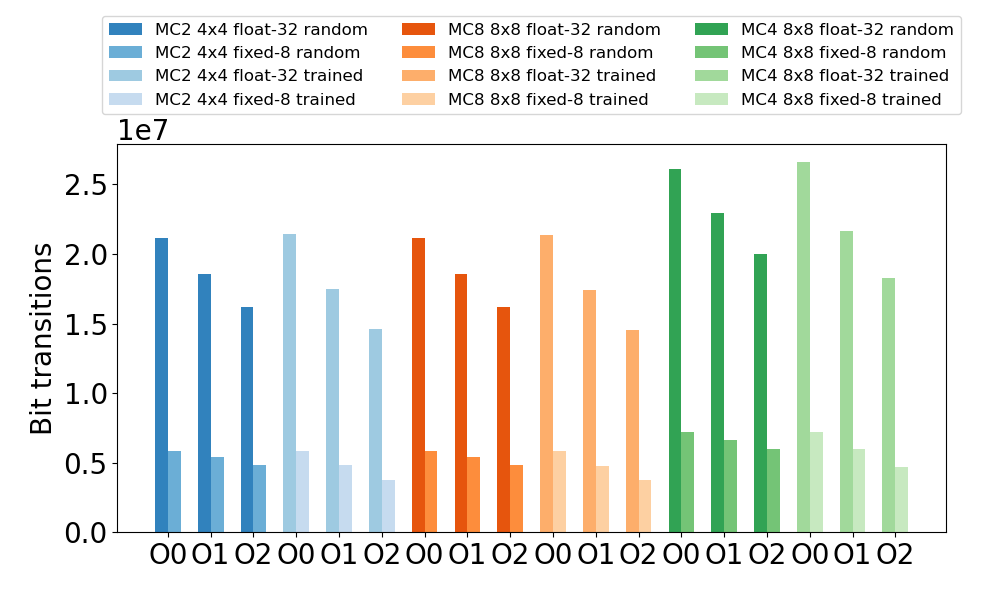} 
	\caption{BTs across different NoC sizes.}
	\label{differenNoCsize.png.}
\end{figure}

 %
For the absolute BT number in \fig{differenNoCsize.png.}, the green MC4 NoC results in more BTs compared to other NoC sizes. The reason is that more routers per MC increase the hops that a flit must traverse, and more hops lead to more overall BTs. 

\par  When comparing BT reduction rates, affiliated-ordering effectively reduces BT from 12.09\% to 18.58\% for floating-32 data and from 7.88\% to 17.75\% for fixed-8 data.  Separated-ordering achieves  BT reduction rate from 23.30\% to 32.01\%  for  float-32 data and from 16.95\% to 35.93\%  for fixed-8 data. A consistent trend emerges: affiliated-ordering achieves significant BT reductions, while separated-ordering consistently delivers the highest reductions.

\subsubsection{Different DNN models}
\par  

\begin{figure}[htbp] %
	\centering 
\includegraphics[trim=0.64cm 0.4cm 0.4cm 0.43cm, clip,width=0.45\textwidth]{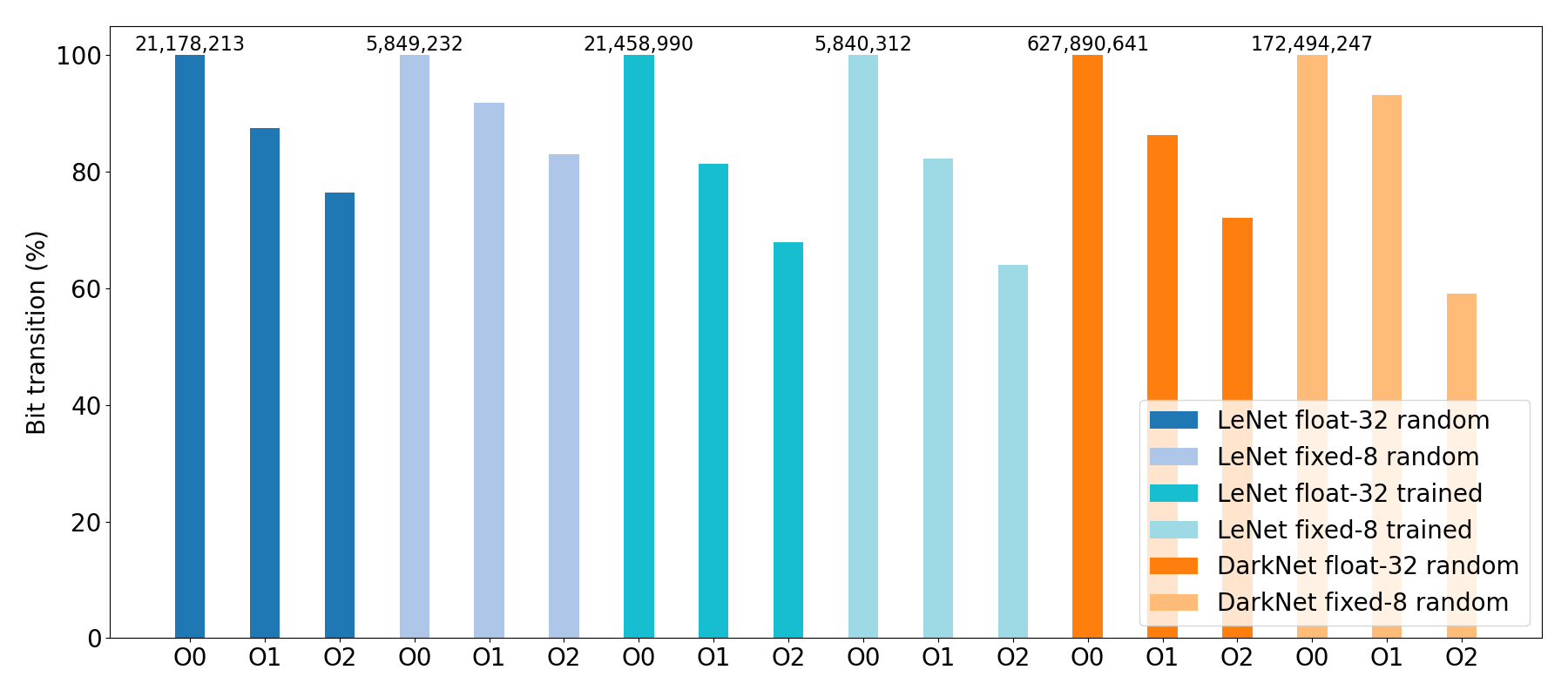} 
	\caption{Normalized BTs for different NN models.}
	\label{differentNNmodels.png.}
\end{figure}

We achieve up to 35.93\% BT reduction for LeNet, and up to 40.85\% for DarkNet in \fig{differentNNmodels.png.}. Aligning with the results across varying NoC sizes, the separated-ordering method achieves the highest BT reduction.

\subsection{Hardware design for ordering unit}
\begin{figure}[htbp]
    \centering
    \begin{minipage}[b]{ 0.48\textwidth}
       \centering 
\includegraphics[ trim=0.64cm 0.4cm 0.4cm 0.63cm, clip, width=\textwidth]{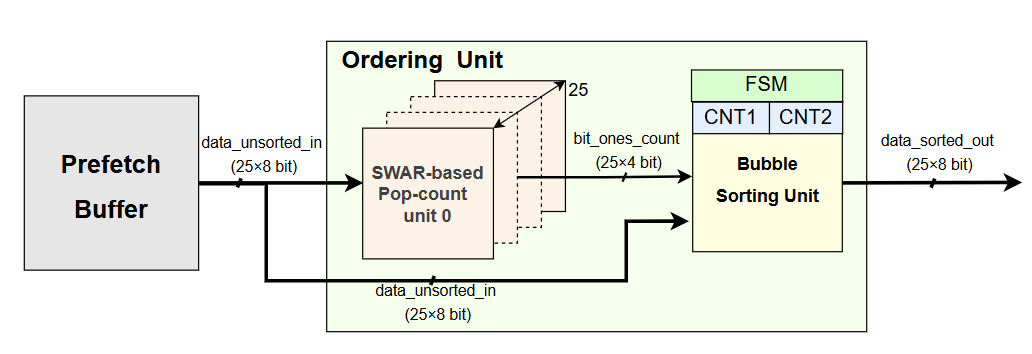} 
	\caption{Hardware design of ordering unit.}
	\label{r2}
    \end{minipage} 
\end{figure}

To evaluate the overhead of our method, we present the hardware design of affiliated-ordering unit. This unit can be used for separated-ordering with double time consumption.
As shown in
 \fig{r2},  classic  Single-instruction-multiple-data Within A Register (SWAR) to count '1'-bits (Pop-count) and bubble sort are implemented.  We synthesize a NoC generated from Constellation \cite{constellationNoC} using Synopsys DC and compare routers with our units in \tab{table_ordering_router}. Our ordering unit occupies only 12.91 kGE (thousand gate equivalents). While a single router consumes 16.92 mW, one ordering unit requires much less power (2.213 mW). The deordering overhead is very light and the count of ordering units equals the number of MCs - for instance, four units in an 8×8 NoC containing 64 routers.
 
\begin{table}[htb]
\centering
\caption{Synthesis Results of Ordering Unit and Router}`
\label{table_ordering_router}
\begin{tabular}{|l|c|c|c|c|}
\hline
\multicolumn{1}{|c|}{Metric} & \multicolumn{2}{c|}{Ordering Unit} & \multicolumn{2}{c|}{Routers} \\ \hline
Technology & \multicolumn{2}{c|}{TSMC 90nm} & \multicolumn{2}{c|}{TSMC 90nm} \\ \hline
Frequency (MHz) & \multicolumn{2}{c|}{125} & \multicolumn{2}{c|}{125} \\ \hline
Voltage (V) & \multicolumn{2}{c|}{1.0} & \multicolumn{2}{c|}{1.0} \\ \hline
Power (mW) & One unit & Four units & One router & 64 routers \\ 
\cline{2-5}
 & 2.213 &  8.852 & 16.92 & 1083.18 \\ \hline
\begin{tabular}[c]{@{}c@{}}Area (Thousand \\ gates  equivalent)\end{tabular}
 & 12.91 & 51.64 & 125.54 & 8034.56 \\ \hline
\end{tabular}
\end{table}

 We synthesize physical links with Innovus and obtain the bit transition energy of  $0.173pJ$. Banerjee \textit{et al.} \cite{banerjeenocs2007power} presents bit transition energy of $0.532pJ$. Assuming half of the 128-bit links transit for an 8x8 NoC with 112 inter-router links, the overall link power under 125MHz is  $ 0.173pJ/bit * \frac{128bits}{2}*112 * 125MHz= 155.008 mW$ for our design and 476.672 mW using Banerjee's \cite{banerjeenocs2007power} link model. The link power and energy depend on many factors, such as physical link length and chip size, material capacitance, etc.  We provide it merely to give an intuitive impression. With the BT reduction of 40.85\% by our method, link power is reduced from 155.008 mW to 91.688 mW or from 476.672 mW to 281.951 mW.

\section{Conclusion}\label{sec:conclusion}
We propose a data transmission ordering approach to reduce BT in NOC-DNA. We present a mathematical analysis showing that a descending ordering based on '1'-bit count is globally optimal for minimizing the BT. Then we evaluate our method with and without NoC. Two ordering methods, affiliated-ordering and separated-ordering, are proposed.   Across diverse configurations—including varying NoC sizes, DNN models, and data formats, our approach achieves significant BT reductions.  We also present the hardware implementation to assess hardware overhead. 

The affiliated-ordering method achieves a BT reduction rate of up to 18.58\% for float-32 data and 17.75\% for fixed-8 data. Separated-ordering effectively reduces BT, achieving up to 32.01\% for float-32 data and  40.85\% reduction in BTs for fixed-8 data. A single ordering unit only occupies 12.91 kGE of area. Four ordering units consume 8.852 mW total power, while 64 routers consume 1083.18 mW.  Combining and comparing this work with other BT reduction works, running more DNN workloads can be explored in the future.

\bibliographystyle{IEEEtran}
\bibliography{ref1}

\end{document}